\documentclass{article}

\usepackage[final]{neurips_2020}

\usepackage[final]{neurips_2020}

\usepackage[utf8]{inputenc} 
\usepackage[T1]{fontenc}    
\usepackage{hyperref}       
\usepackage{url}            
\usepackage{booktabs}       
\usepackage{amsfonts}       
\usepackage{nicefrac}       
\usepackage{microtype}      
\usepackage{wrapfig}
\usepackage{multirow}
\usepackage{graphicx}
\graphicspath{{./Figures/}}

\title{Investigating two super-resolution methods for downscaling precipitation: ESRGAN and CAR}

\author{%
  Campbell D. Watson \\
  IBM Research, New York, USA\\  
  \texttt{cwatson@us.ibm.com} \\
  \And
  Chulin Wang \\
  IBM Research, New York, USA \\
  Northwestern University, IL, USA \\
  \texttt{wangc@u.northwestern.edu} \\
  \And
  Timothy Lynar \\
  University of New South Wales, Canberra, Australia\\
  \texttt{t.lynar@adfa.edu.au} \\
  \And
  Komminist Weldemariam \\
  IBM Research, Nairobi, Kenya\\
  \texttt{k.weldemariam@ke.ibm.com} \\
}

\begin{document}

\maketitle

\begin{abstract}
  In an effort to provide optimal inputs to downstream modeling systems (e.g., a hydrodynamics model that simulates the water circulation of a lake), we hereby strive to enhance resolution of precipitation fields from a weather model by up to 9x. We test two super-resolution models: the enhanced super-resolution generative adversarial networks (ESRGAN) proposed in 2017, and the content adaptive resampler (CAR) proposed in 2020. Both models outperform simple bicubic interpolation, with the ESRGAN exceeding expectations for accuracy. We make several proposals for extending the work to ensure it can be a useful tool for quantifying the impact of climate change on local ecosystems while removing reliance on energy-intensive, high-resolution weather model simulations.
\end{abstract}

\section{Introduction}

Weather prediction at a high spatial resolution is valuable and often essential for many applications including flood prediction and water management, renewable energy production, and wildfire forecasting. However, its production using traditional approaches (e.g., numerical weather prediction models) is computationally expensive. Recently, researchers have been exploring the use of machine learning techniques to downscale coarser resolution weather and climate simulations to finer resolution grids. Broadly speaking, four machine learning methods for have been described in the literature: Regression based methods, analog methods, auto-regression methods, and deep neural network super resolution based methods (e.g., \cite{Larraondo2017, Mouatadid2017, Coulibaly2005, He2016, Rodrigues2018, Medina2018, Rebora2006}. The latter method encompasses convolutional and generative adversarial neural networks for super resolution.

The motivation for this work stems from The Jefferson Project at Lake George, a multiyear research initiative based at Lake George, New York, USA that has a goal of understanding the impact of human activity on fresh water, and how to mitigate those effects. A recent discovery (see \cite{AugerSubmitted}) is that hydrodynamic simulations of Lake George are considerably more accurate when forced with high-resolution weather model data (<1 km horizontal resolution). A particularly striking result is that after one month of simulation, the hypolimnetic volume of the lake reduces by 15\% when forced with weather information generated at 0.33 km resolution -- in line with observations -- compared to 3 km. In other words, the livable zone for cold-water fish reduces by 15\%, which has significant implications for understanding the impact of a warming climate on lake biological activity.

Given the apparent importance of high-resolution weather data to accurately model the long-term changes of mid- and small-sized lakes, and given the exorbitant computational expense (and carbon footprint) of generating such high-resolution simulations with a physics-based weather model, we hereby examine the feasibility of using two super-resolution models to generate high-resolution inputs from coarse resolution weather model data. We strive for a resolution increase of up to 9x, more than most recent ML-based downscaling studies (e.g., \cite{Manepalli2020}). Successful implementation of such algorithms will enable the generation of high-resolution weather information at significantly lower computational cost. This, in turn, may spur a deeper, global understanding of local weather impacts on vulnerable ecosystems in a changing climate.

The manuscript presents early results for precipitation only, although other weather variables (wind, downward radiation, humidity) are equally important for coupling with hydrodynamic simulations. It builds on previous attempts at downscaling using tree-based regression models trained at every grid cell (\cite{WatsonLynar2020}), which were found to be inferior to super-resolution algorithms.

\section{Data and Methods}

\begin{wrapfigure}{r}{0.5\textwidth}
 \begin{center}
 \includegraphics[width=0.48\textwidth]{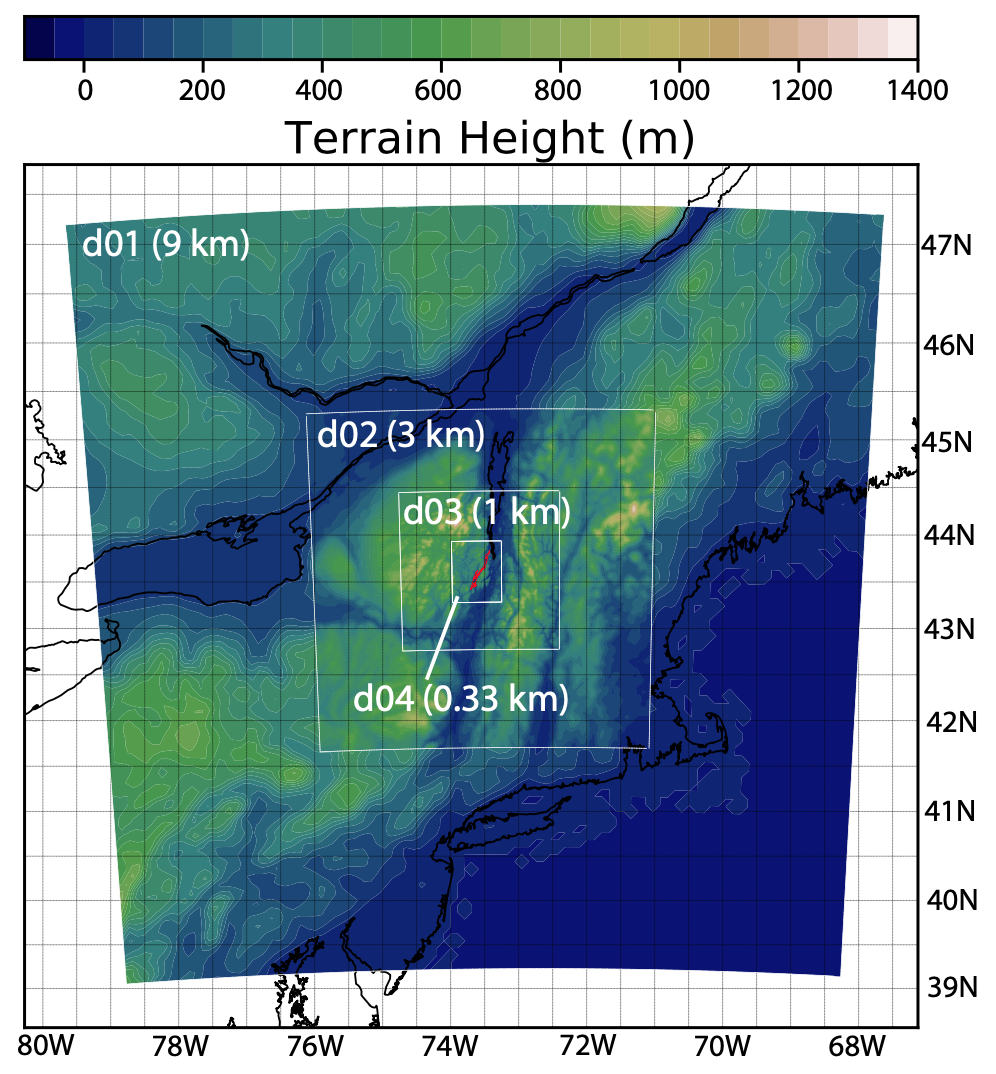}
 \end{center}
 \caption{WRF model domains in northeast USA (Lake George is outlined in red). This manuscript uses WRF data from the three outer domains with 9, 3 and 1 km grid resolution.}
 \label{fig:Domains}
\end{wrapfigure}

All precipitation data used in this study was generated by version 3.8.1 of the Advanced Research version of the Weather Research and Forecasting (WRF) Model (\cite{Skamarock08}). Daily, operational weather forecasts for Lake George have been performed for Lake George with four (one-way coupled) domains with horizontal resolutions of 9 km, 3 km, 1 km and 0.33 km (Fig. \ref{fig:Domains}). Each domain has 42 vertical levels (with 14 below 1 km), a 50-hPa model top, and a 5 km-deep upper-level absorbing layer. Further details can be found in \citet{AugerSubmitted}.

We use hourly WRF data from 2017-01-01 to 2020-03-20 (with an 80\%/20\% split at 2019-08-01) in an effort to reconstruct the simulated 1 km WRF data with 3 km and 9 km WRF data. In other words:

\begin{itemize}

\item Experiment 1: Feature values are 3 km WRF variables; target variable is 1 km WRF precipitation 

\item Experiment 2: Feature values are 9 km WRF variables; target variable is 1 km WRF precipitation

\end{itemize}

The WRF variables used in this study are hourly accumulated surface precipitation, and instantaneous 2 m temperature, u- and v- 10 m winds, 2 m specific humidity and surface pressure. All variables are output at hourly intervals.

\subsection{Downscaling with ESRGAN}

The enhanced super-resolution generative adversarial networks (ESRGAN) is a deep neural network approach which was initially created for use with natural images which typically have no inherent resolution. Image super-resolution models have received increased application in the physical sciences, with \citet{Manepalli2020} finding the ESRGAN has encouraging transferable and generalizable properties (and can even reproduce the simulated power spectral density of near-surface winds).

We employed the SRResNet structure with the Residual-in-Residual Dense Block (RRDB) as basic blocks, as proposed by \citet{Wang2019} paper. We used the default training parameters mentioned in ESRGAN, except with 3x- instead of 2x-upsampling blocks. In order to optimize for the prediction accuracy rather than visual quality, the training is PSNR-oriented without adding VGG or GAN loss (as per the recommendation in \citet{Wang2019}).  

\subsection{Downscaling with Content Adaptive Resampler}

Recently, \citet{sun2020} proposed a new content adaptive-resampler (CAR) based image downscaling method. Briefly, the resampler network generates content adaptive image resampling kernels which are applied to the original high-resolution input to generate pixels on the downscaled image. This resampler is different to traditional bilinear downscaling as it is dynamic and uses meta-learning that in turn utilises filters similar to dynamic filter networks. We employed the CAR model implemented with Pytorch using the default configurations as proposed by \citet{sun2020}. 

\section{Preliminary Results}

The ESRGAN systematically outperformed the CAR model, presenting accurate reconstructions well beyond expectations. The long timeseries of WRF data ensures ample training and testing examples, with the 80/20\% train/test split providing almost 8 months of continuous data for testing the skill of the models. Overall, the mean absolute error (MAE) is shown in Table~\ref{tab:MAE} along with that obtained via standard bicubic interpolation (a simple baseline model).

\begin{table}[h]
    \centering
    \caption{Mean absolute error using different methods. The output grid resolution is 1 km, and the input grid resolution is listed in the table. The ESRGAN model has the best performance, followed by the CAR model. Both of the proposed models outperform the baseline bicubic interpolation.}
    \begin{tabular}{cccc}
        \toprule
          \multirow{2}{*}{Input grid resolution} & \multicolumn{3}{c}{Mean absolute error (mm/hr)} \\
          \cmidrule(r){2-4}
         & Bicubic & ESRGAN & CAR \\
        \midrule
        3 km & 0.83 & 0.040 & 0.384\\
        9 km & 1.98 & 0.078 & 0.931\\
        \bottomrule
    \end{tabular}
    \label{tab:MAE}
\end{table}

\begin{figure}[h]
  \centering
  \includegraphics[width=\textwidth]{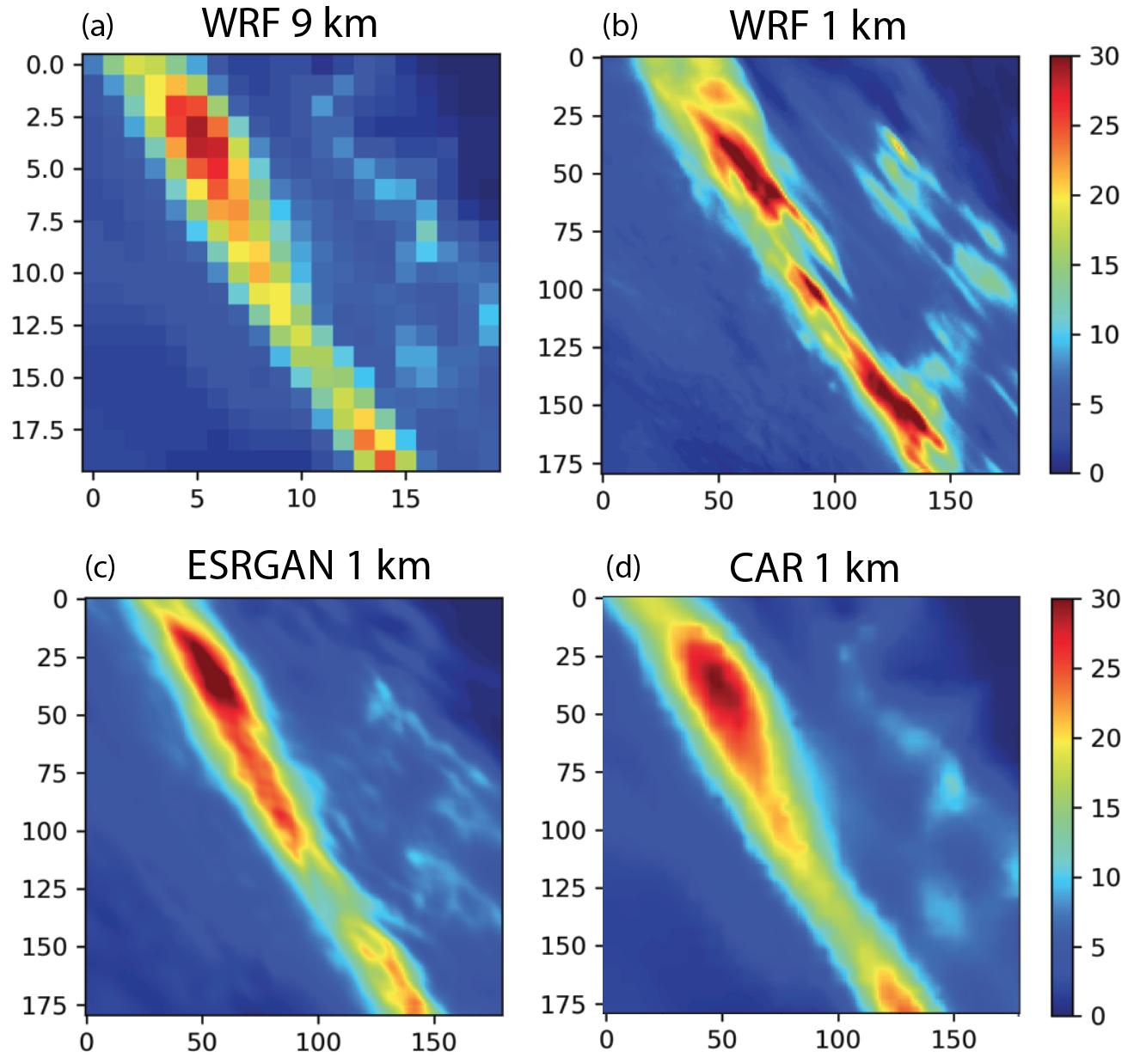}
  \caption{Example of precipitation reconstructions by the ESRGAN and CAR models from 9 km to 1 km horizontal resolution. The panels show accumulated precipitation (mm/hr) in the hour preceding 2019-11-01 04:00 UTC. (a) Original WRF precipitation from the 9 km domain; (b) original WRF precipitation from the 1 km domain (target); reconstruction by ESRGAN to 1 km resolution; and (d) reconstruction by the CAR model to 1 km resolution}
  \label{fig:4panel}
\end{figure}

Specific events were extremely well captured by the ESRGAN (e.g., Fig. \ref{fig:4panel}), and although it did systematically underestimate precipitation by up to 0.1 mm/hr (not shown), it motivates us to continue investigation with this model. In contrast, the CAR method retained many visual features of the coarser resolution precipitation input; perhaps additional testing with different configurations can improve results. 
 
\section{Discussion and Broader Impacts}

This study has demonstrated the utility of two super-resolution methods to downscale precipitation in the northeast US. The ESRGAN outperformed the CAR method, especially when downscaling from 9 km to 1 km precipitation. The significant 9x super-resolution is particularly encouraging because the WRF model uses different physics to simulate precipitation at 9 km resolution vs 1 km (we used the Grell-Freitas cumulus scheme in the 9 km domain, whereas cumulus-scale precipitation is assumed to be explicitly simulated in the 1 km domain). 

We are excited for the potential impact of this work. An accurate and efficient downscaling method reduces time spent running expensive computational code (and thus reduce the carbon footprint) and provides a scalable approach for (global) climate impact assessments.

Further investigation has been identified:

\begin{itemize}

\item Expand analysis to include additional variables, specifically near surface winds, surface humidity and downward short and long wave radiation. These are key inputs to hydrodynamic models.

\item Apply ESRGAN to downscale GFS data (which provides boundary conditions to the WRF model) at ~25 km resolution to the 1 km WRF domain. This is a particularly demanding task given the GFS model has different model parameterizations and dynamical core.

\item How transferable and generalizable is this approach? \citet{Manepalli2020} showed some success in this regard using ESRGAN to downscale winds, and is key to moving to climatic geographies. 

\item How much data is required to adequately train the model? Here, we provided over nearly 2.5 years of hourly data for training; a prohibitive amount of simulation for larger geographic areas. Can we leverage any correlations across time to improve the super-resolution (e.g., low resolution sequence in, high resolution sequence out)?

\end{itemize}

Finally, we do not believe the data is biased nor do we expect people to be placed at a disadvantage from this research. That said, flaws in the super-resolution system could result in inaccurate weather and climate predictions with net-negative consequences for policy development and decision making.

\begin{ack}

We gratefully acknowledge assistance by personnel associated with the Jefferson Project at Lake George, a collaboration between Rensselaer Polytechnic Institute, IBM Research and the FUND for Lake George. In addition to funding from the three partners, the sensor network has been partially supported by a NSF MRI grant (\#1655168) and a New York State Higher Education Capital grant (\#7290).

\end{ack}

\medskip

\small

\bibliography{Biblio}

\end{document}